\documentclass[prl,reprint]{revtex4-1}

\usepackage{hyperref}
\usepackage{amsmath}
\usepackage{amsfonts}
\usepackage{amssymb}
\usepackage{epsfig}
\usepackage{graphicx}

\begin{document}

\title{Optical Signatures of Dirac Nodal-lines in NbAs$_2$}

\author{Yinming Shao$^{1,2,}$}
\email{ys2956@columbia.edu}
\author{Zhiyuan Sun$^{2}$, Ying Wang$^{3}$, Chenchao Xu$^{4}$, R. Sankar$^{5}$, Alex J. Breindel$^{2}$,\\Chao Cao$^{7}$,
 M. M. Fogler$^{2}$, Fangcheng Chou$^{5}$, Zhiqiang Li$^{6}$, T. Timusk$^{8,9}$, M. Brian Maple$^{2}$ and D. N. Basov$^{1,2}$}

\affiliation
{$^{1}$Department of Physics, Columbia University, New York, New York 10027, USA\\
$^{2}$Physics Department, University of California-San Diego, La Jolla, California 92093, USA\\
$^{3}$National High Magnetic Field Laboratory, Tallahassee, Florida 32310, USA\\
$^{4}$Department of Physics, Zhejiang University, Hangzhou 310036, China\\
$^{5}$Center for Condensed Matter Sciences, National Taiwan University, Taipei 10617, Taiwan\\
$^{6}$College of Physical Science and Technology, Sichuan University, Chengdu, Sichuan 610064, China\\
$^{7}$Department of Physics, Hangzhou Normal University, Hangzhou 310036, China\\
$^{8}$Department of Physics and Astronomy, McMaster University, Hamilton, Ontario L8S 4M1, Canada\\
$^{9}$The Canadian Institute for Advanced Research, Toronto, Ontario M5G 1Z8, Canada
}

\begin{abstract}
Using polarized optical and magneto-optical spectroscopy, we have demonstrated universal aspects of electrodynamics associated with Dirac nodal-lines. We investigated anisotropic electrodynamics of NbAs$_2$ where the spin-orbit interaction triggers energy gaps along the nodal-lines, which manifest as sharp steps in the optical conductivity spectra. We show experimentally and theoretically that shifted 2D Dirac nodal-lines feature linear scaling $\sigma_1 (\omega)\sim\omega$, similar to 3D nodal-points. Massive Dirac nature of the nodal-lines are confirmed by magneto-optical data, which may also be indicative of theoretically predicted surface states. Optical data also offer a natural explanation for the giant magneto-resistance in NbAs$_2$.
\end{abstract}

\maketitle

Nodal-line semimetals (NLSM) are newly discovered quantum materials with linear bands and symmetry protected band degeneracies. Compared to three-dimensional (3D) Dirac/Weyl semimetals, the band-touchings in NLSMs are not constrained to discrete points but extend along lines in the Brillouin zone (BZ) \cite{burkov2011,fang2015,weng2016,armitage2018a}. NLSMs can be considered as precursors to many other topological phases \cite{weng2016}, including Weyl semimetals \cite{huang2015a,weng2015}. Despite intense interests \cite{liu2017h,yan2016,lim2017,nie2017a} and numerous material predictions \cite{weng2016,armitage2018a}, the experimental identification of NLSMs has been rare \cite{bian2016,schoop2016}, limited by the surface-sensitive nature of available probes. On the other hand, nontrivial topologies in quantum materials are often revealed via nontrivial response functions \cite{armitage2018a}. Bulk probes \cite{hu2016,singha2017a,schilling2017b} and characteristic response functions \cite{carbotte2017,mukherjee2017a,ahn2017b} are therefore the key to identify NLSMs and investigate the complex bulk nodal-lines.

Power law behavior of the real part of the optical conductivity ($\sigma(\omega)\!=\!\sigma_1(\omega)+i\sigma_2(\omega)$) over extended frequency, i.e., $\sigma_1(\omega)\!\sim\!\omega^{d-2}$ \cite{hosur2012,bacsi2013,tabert2016}, is a hallmark of Dirac-like nodal-points in solids. Linear ($\sigma_1(\omega)\!\sim\!\omega$) and constant optical conductivity has been confirmed in 3D (e.g., pyrochlore iridates \cite{sushkov2015}, Dirac semimetal Cd$_3$As$_2$ \cite{neubauer2016a}, ZrTe$_5$ \cite{chen2017}) and 2D (e.g., graphene \cite{li2008}), respectively. Here we show experimentally and theoretically that \textit{energy-shifted} 2D Dirac nodal-lines can also give rise to linear optical conductivity. Similar to other Dirac materials \cite{li2008,neubauer2016a}, the power law of $\sigma_1(\omega)$ breaks down below the gap energy. We refer to these gapped nodal-lines simply as nodal-lines and the node is understood as the Dirac point of the massive Dirac band \cite{schilling2017b,patri2017}. A particularly interesting effect pertains to gapping of the nodal-lines as the result of spin-orbit coupling (SOC) \cite{fang2015,armitage2018a,schilling2017b}: a phenomenon we are set to investigate in NbAs$_2$ single crystals. The magnitude of the gap is linearly proportional to the strength of SOC, which follows the order Nb$<$Ta and P$<$As$<$Sb in transition metal dipnictides \cite{xu2016a}.

We explore the electrodynamics of shifted or dispersive nodal lines using NbAs$_2$ as a case study. Importantly, the observed steps and linear power law in the optical conductivity reflect the gaps and the energy-dispersions of the nodal-lines. Furthermore, the Dirac linear dispersion perpendicular to the lines is established by $\sqrt{B}$-spaced Landau-levels (LLs) in magneto-optics, previously observed only for nodal-points \cite{orlita2011,shao2017,shuvaev2017}. The nodal-lines discovered through unusual response functions here also naturally explain the exotic magnetoresistance (MR) properties of NbAs$_2$ \cite{shen2016,yuan2016a,li2016a}.

\begin{figure}[ht]
\includegraphics[width=0.9\columnwidth]{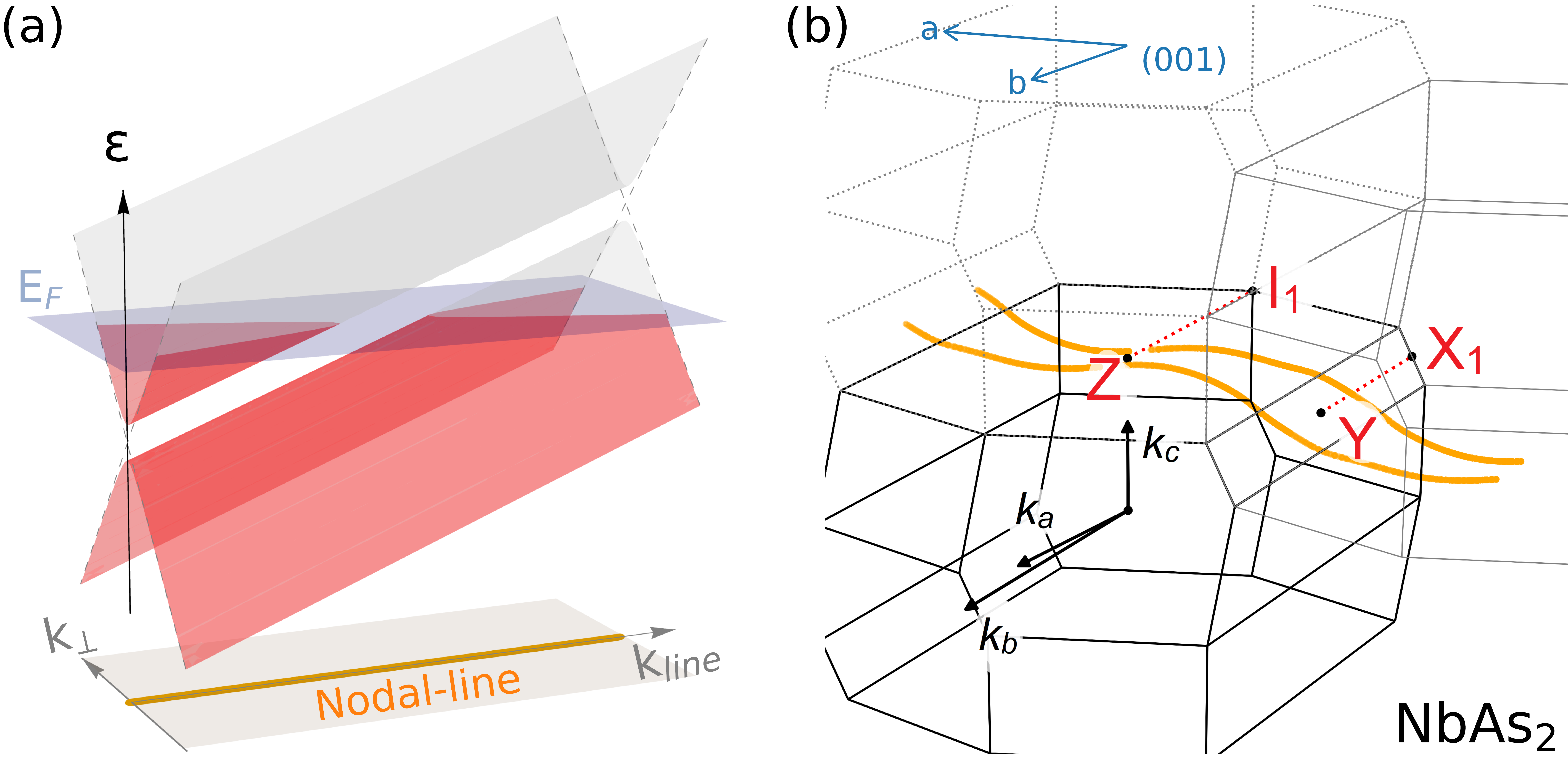}
\caption{(a) Energy versus momentum dispersion for a energy-shifted nodal-line. Purple and red color indicate the Fermi energy plane and filled bands, respectively. The orange line is the nodal-line projected in momentum space. (b) $ab$-$inito$ calculations of the nodal-lines (orange) in momentum space for NbAs$_2$. Red symbols are the high symmetry points in the Brillouin zone near the nodal-lines.}
\label{fig1}
\end{figure}

\begin{figure*}[!ht]
\includegraphics[width=1\textwidth]{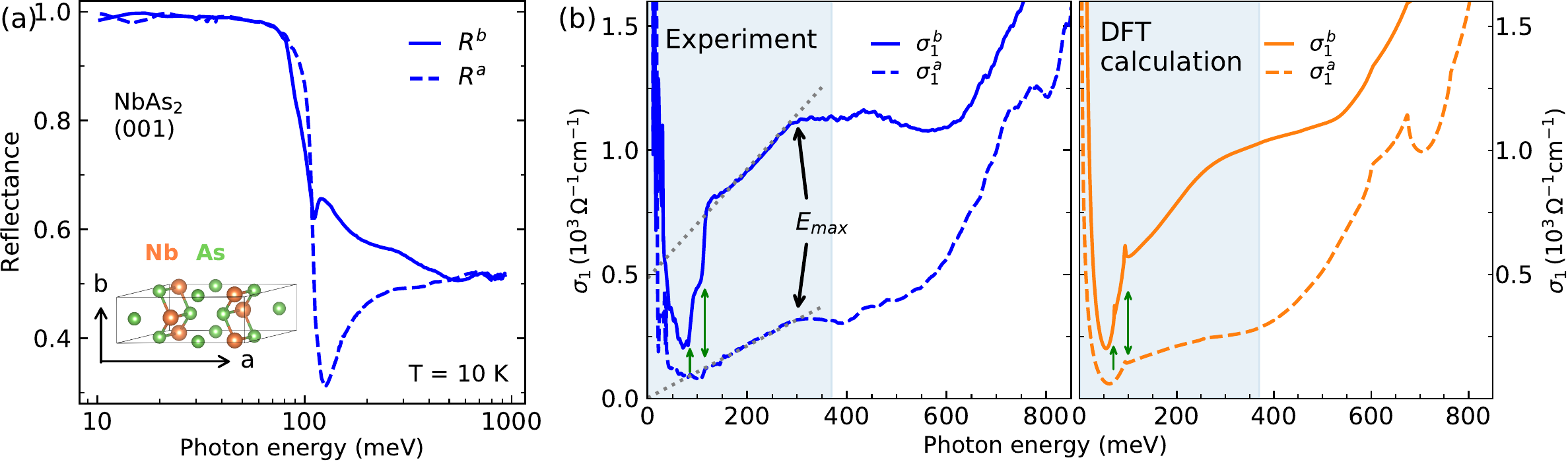}
\caption{(a) Anisotropic reflectance on the NbAs$_2$ (001) surface. Top inset is the primitive BZ, and the orange lines are calculated nodal-lines (without SOC). Bottom inset is a schematic of one unit cell of NbAs$_2$. (b) Optical conductivity from experiment (left) and DFT calculations (right). Blue shaded regions highlight the low-energy part where the response is dominated by the massive Dirac bands. Green arrows indicate step positions in $\sigma_1(\omega)$. In all panels, solid and dashed lines indicate crystallographic $b$-axis and $a$-axis response, respectively.}
\label{fig2}
\end{figure*}

Fig.~\ref{fig1}(a) shows a schematic of the shifted nodal-line in energy-momentum space where the Fermi energy (E$_F$) plane passes through the line. The projected nodal-line in momentum space is also shown at the bottom. Realistic NLSM materials often have complicated nodal-line structure in momentum space. Fig.~\ref{fig1}(b) is the $\textit{ab-inito}$ calculations of the nodal-lines (orange) in NbAs$_2$ using density functional theory (DFT). Contrary to most nodal-line models (nodal-rings \cite{carbotte2017,mukherjee2017a,ahn2017b}), the nodal-lines in NbAs$_2$ are open-ended and extends indefinitely through multiple BZs. The directionality of the open nodal-lines implies huge optical anisotropy since the $\sigma_1(\omega)$ is predicted to vanish along the nodal-line direction \cite{carbotte2017,mukherjee2017a,ahn2017b}. 

The polarized reflectance spectra of the NbAs$_2$ (001) surface at 10 K is shown in Fig. \ref{fig2}(a). The $a$-axis reflectance (R$^a$) shows a pronounced plasma minimum ($\sim$125 meV) near the screened plasma frequency. For the $b$-axis reflectance (R$^b$), the plasma edge appears broadened and a sharp dip develops around 110 meV. While the far-infrared reflectance for both polarizations show similar metallic behavior below 50 meV (R$\,\sim\,$1), the mid-infrared response is highly anisotropic. 

In Fig. \ref{fig2}(b), left panel, we display the 10 K optical conductivity for both polarizations of light. The Drude conductivity in both $\sigma_1^a$ and $\sigma_1^b$ feature at least two free-carrier components, consistent with multiple Fermi pockets revealed by quantum oscillation measurements in NbAs$_2$ \cite{shen2016,yuan2016a,li2016a}. The most striking feature is the sharp double-step in $\sigma_1^b$ (green arrows), followed by $\sigma_1 (\omega)\!\sim\!\omega$ relation over an extended frequency range. Weaker step structure and linear conductivity are also evident in $\sigma_1^a$. Interestingly, the double-step structure and linear conductivity above it resemble the predicted $\sigma_1 (\omega)$ for an inversion-symmetry breaking Weyl semimetal \cite{tabert2016} (e.g., NbAs). The lack of inversion symmetry breaks the degeneracy of the two Weyl cones and causes them to shift in energy \cite{tabert2016}. The Pauli exclusion principle dictates that an optical transition will be forbidden if the final states are filled (Pauli-blocking). In the energy-shifted cones, Pauli-blocking happens at different energies, hence the predicted double-step appear. Although no Weyl points exist in NbAs$_2$, the nodal-lines give rise to linearly growing $\sigma_1 (\omega)$ above the gaps, which we will focus on next. 
\begin{figure}[!h]
\includegraphics[width=1\columnwidth]{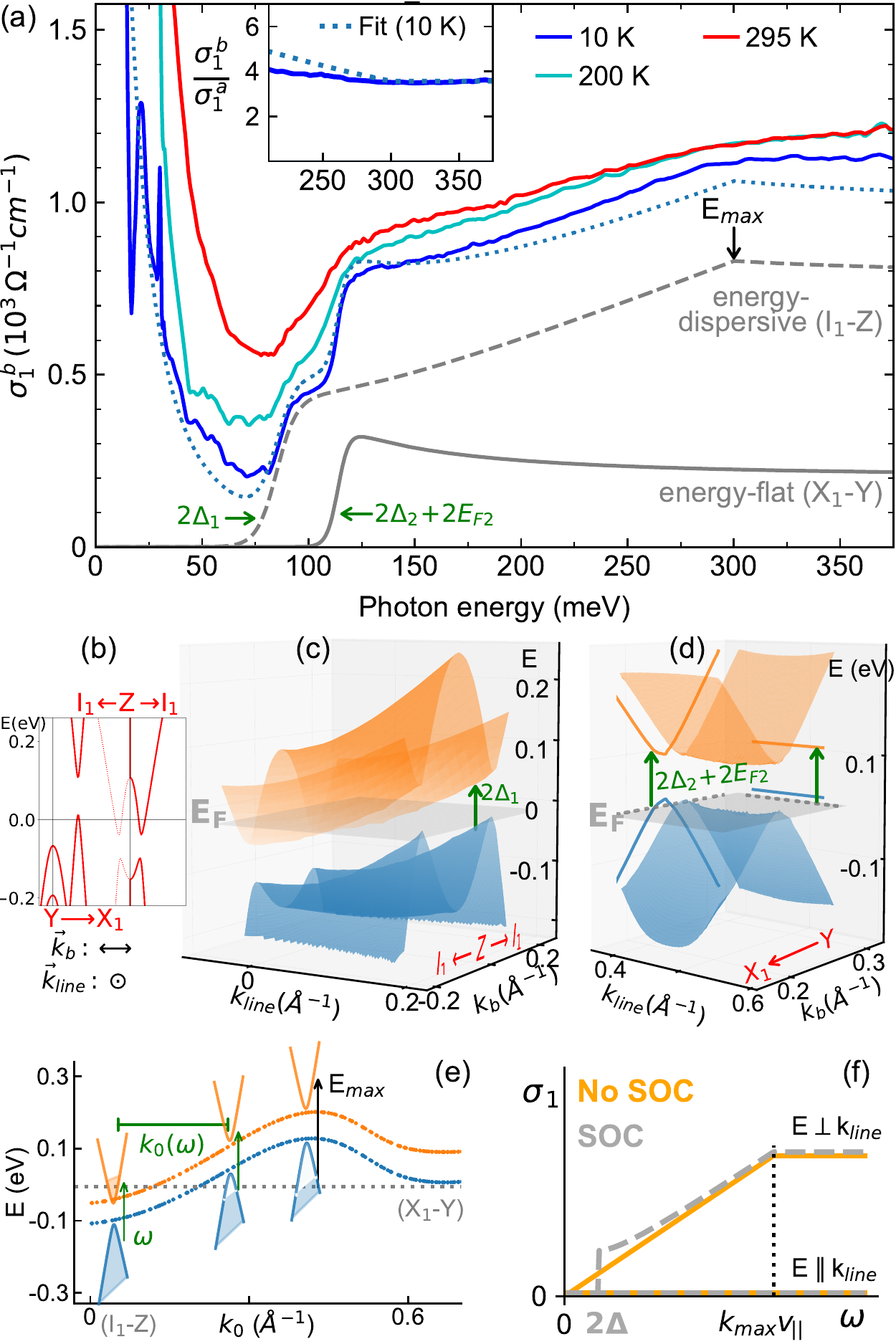}
\caption{(a) Optical conductivity for E$\,||\,b$. Blue dotted lines are fitted $\sigma_1^{b}$ curve with nodal-line structure parameters. Grey dashed and solid lines denote contributions from the nodal-lines near I$_1$-Z (panel (c)) and X$_1$-Y (panel (d)), respectively. The linear increase saturates at $E_{max}\!\sim$0.3 eV. The inset shows the ratio $\sigma_1^{b}$/$\sigma_1^{a}$ above the gap region. Panel (b) displays the band structure calculation along high symmetry points near the nodal-line regions and its 3D version is shown in (c) and (d). Green arrows illustrate onsets of interband transitions for dispersive (c) and energy-flat (d) part of the nodal-line. (e) Calculated nodal-line energy vs. line length $k_0$. Dirac-cone schematics indicate different fillings of the Dirac bands along the line. Grey dotted line is the Fermi energy. Vertical arrows show different onsets of interband transition and horizontal arrow is the effective line length. (f) corresponding optical conductivity for the dispersive region in (e).}
\label{fig3}
\end{figure}

Optical conductivities calculated using DFT are shown in Fig. \ref{fig2}(b), right panel. The DFT calculation captures the gross features of the data, including the steps, the linear dependence and the slope change at $E_{max}\!\sim$0.3 eV in both $\sigma_1^{a}$ and $\sigma_1^{b}$. The anisotropy between $\sigma_1^{a}$ and $\sigma_1^{b}$ is also evident from the calculations. While the overall agreement is good, certain intriguing features are not readily understood in the DFT calculation. Importantly, while the linear slope extrapolates close to 0 at zero energy for $\sigma_1^{a}$, both the experiment and the calculations show a large, non-zero intercept for $\sigma_1^{b}$. This large intercept at zero energy is inconsistent with the optical conductivity model for 3D Dirac/Weyl fermions mentioned before. Instead, we show that the linear conductivity and the intercept result from the nodal-line in NbAs$_2$. SOC triggers energy gaps along the nodal-line (Fig. \ref{fig3}(e)) and the gap size changes from $\sim100$ meV (2$\Delta_2$) near the high symmetry line X$_1$-Y to $\sim80$ meV (2$\Delta_1$) near I$_1$-Z. Both X$_1$-Y and I$_1$-Z are parallel to the $k_b$ direction (see Figs. \ref{fig3}, S2). We demonstrate below that while a flat nodal-line (near X$_1$-Y) gives rise to constant $\sigma_1(\omega)$, the shifted nodal-line near I$_1$-Z leads to linear conductivity in NbAs$_2$. The combination of dispersive and flat nodal-lines causes the observed linear optical conductivity with a finite intercept.

A shifted nodal-line is described by the band dispersion
\begin{equation} \label{eqn:nodal_band}
\varepsilon_{\pm} = \pm \sqrt{\Delta^2 + v_1^2 k_1^2 + v_2^2 k_2^2} + v_\parallel k_\parallel
\end{equation}
where $k_\parallel$ is the momentum along the nodal line while $k_1$ and $k_2$ are those perpendicular to it. As shown in Fig.~\ref{fig1}(a), there is an overall energy-shift along the nodal line quantified by the `shifting' velocity $v_\parallel$. Perpendicular to the nodal line, the dispersion is Dirac-like with the asymptotic velocities $v_1$ and $v_2$. Spin-orbit coupling induces a finite mass $\Delta$.
For a general (massive) nodal-line, we have derived the corresponding real optical conductivity as follows (Supplementary Information (SI) Sec. II): 
\begin{equation} \label{eq1}
    \sigma_{NL}^i (\omega) = \frac{N}{16}\frac{e^2}{h}k_0(\omega)\frac{v_{i}^2}{v_1 v_2} \left(\!1+\frac{4\Delta^2}{\omega^2}\right)\!\Theta(\omega-2\Delta_{op})
\end{equation}
where $v_i$ is the asymptotic velocity along the electric field direction. Note that along the nodal-line direction $v_3$=0 and the corresponding $\sigma_{NL}$ vanishes. $N$ is the degeneracy of nodal-lines, $e$ is electric charge, $h$ is Planck's constant, $k_0(\omega)$ is the effective nodal-line length in $\mathbf{k}$-space where optical transition actually takes place, $\Theta$ is a step function and $2\Delta_{op}$ is the optical gap ($2\Delta+2E_F$). If the nodal-line length is independent of energy ($k_0(\omega)$=$k_0$), the simple flat optical conductivity $\sigma_{NL}(\omega)\!\sim\!\frac{e^2}{h}k_0$ occurs above the gap. However, once energy-dependent nodal-line length is considered, the optical conductivity attains the same frequency dependence as $k_0(\omega)$. Therefore, $\sigma_{NL}(\omega)$ provides direct access to the complex geometry of a nodal-line in $\mathbf{k}$-space via its length $k_0(\omega)$. For linearly shifted nodal line as described by Eq. \ref{eqn:nodal_band}, $k_0(\omega)$=$\omega$/$v_{\parallel}$ and the interband optical conductivity becomes $\sigma_{1}(\omega)\!\sim\!\omega$.

In Fig. \ref{fig3}(a), $\sigma_1^{b}(\omega)$ data at three different temperatures are shown, highlighting the broadening of steps at higher temperatures. The blue dotted line is the fitted total $\sigma_1^b$ (10 K), showing excellent agreement with experiment. Grey dashed and solid lines are the fitted interband contributions using Eq. \ref{eq1}. The fitting parameters are listed in Table S1. The same parameters produce fitted $\sigma_1^a$ curve that are in excellent agreement with experiment as well (see Fig. S3). In Figs. \ref{fig3}(b)-(d), we plot the calculated band structure near I$_1$-Z and near X$_1$-Y for momentum directions $k_{line}$ and $k_b$ ($k_{line}\perp k_b$). The grey planes indicate constant Fermi energy ($E_F$). The side walls of Fig. \ref{fig3}(d) show the projected band structure along each direction, highlighting the anisotropy in asymptotic velocities of nodal-lines. The nodal-lines in \ref{fig3}(c) and \ref{fig3}(d) belong to the same nodal-line, but feature near-linearly dispersing (near I$_1$-Z) and flat (near X$_1$-Y) regions with respect to line length $k_0$ (Fig. \ref{fig3}(e)). 
\begin{figure*}[ht]
\includegraphics[width=0.98\textwidth]{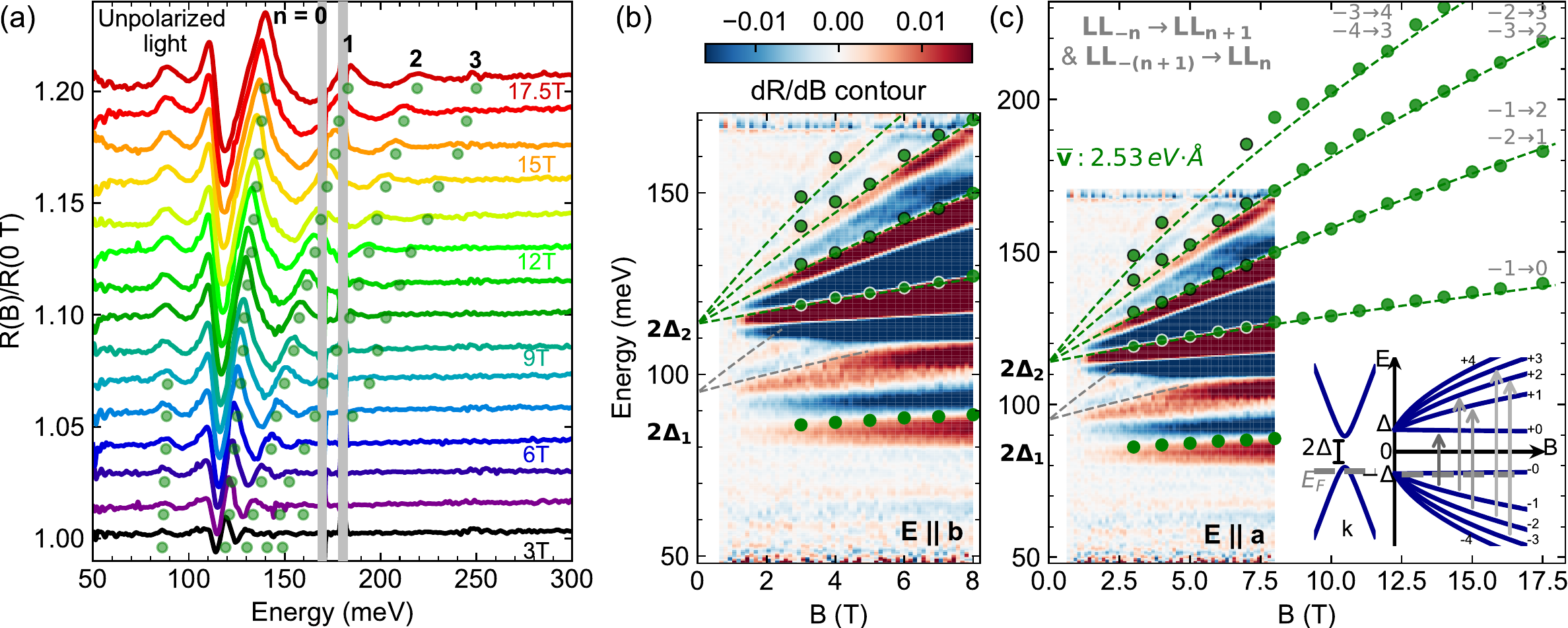}
\caption{(a) Magneto-reflectance spectra normalized by zero-field reflectance, showing a series of LL transitions moving with $B$ and a weakly dispersing mode at $\sim$85 meV. (b) Derivative contour ($d$R/$d$B) for E$\,||\,b$. Dispersive peak energies extracted from panel (a) are overlaid on top of the contour as green dots. Grey dashed lines indicate the dispersive in-gap states. (c) $d$R/$d$B for E$\,||\,a$ and peak energies extracted from A. Green dashed lines in (b) and (c) are fits using Eq. \ref{eq3} with the same parameters. Inset of C shows gapped Dirac bands and the LL dispersion with magnetic field $B$. Arrows indicate allowed interband LL transitions across the gap. }
\label{fig4}
\end{figure*}

An intuitive picture for the linear law of the optical conductivity from extended Pauli-blocking is presented in Figs. \ref{fig3}(e) and \ref{fig3}(f). Orange and blue dotted line in Fig. \ref{fig3}(e) indicate the calculated gap energies at different line lengths $k_0$ (from I$_1$-Z to X$_1$-Y). Schematics of Dirac cones are overlaid on the calculation to illustrate the filling level change along the line. Green vertical arrows indicate the onset of interband transition as a result of Pauli-blocking. With increasing photon energy ($\omega_2\!>\!\omega_1$), longer length of the nodal-line ($k_0(\omega_2)\!>\!k_0(\omega_1)$) are activated as Pauli-blocking is extended to larger phase space. The resulting $\sigma_1(\omega)$ grows linearly until the entire dispersive nodal-line ($k_{max}$) is activated ($E_{max}\!\sim\!k_{max}v_{\parallel}$). For a gapless nodal-line the linear power law of $\sigma_1(\omega)$ extrapolates to zero (Fig. \ref{fig3}(f)). This simple picture captures all the features in $\sigma_1$ for the dispersive nodal-line (I$_1$-Z).

While the step in $\sigma_1$ is pinned to $2\Delta_1$ for the dispersive nodal-line crossing $E_F$ (Fig. \ref{fig3}(c)), the step associated with the energy-flat nodal-line occurs at $2\Delta_2$+$2E_{F2}$ (Fig. \ref{fig3}(d)). The constant optical conductivity causing the finite intercept for $\sigma_1^{b}$ is absent in $\sigma_1^{a}$ since the nodal-line is nearly parallel to $a$-axis (SI Sec. II.C). According to Eq. \ref{eq1}, the anisotropy of the conductivity should also be frequency-independent above the gap energy ($\sigma_1^{b}$/$\sigma_1^{a}\!\sim\!v_b^2$/$v_a^2$), in agreement with experiment (inset of Fig. \ref{fig3}(a)). The anisotropic optical conductivities therefore demonstrate not only the existence of both flat and dispersive nodal-lines in NbAs$_2$, but also the suppression of conductivity along the nodal-line. We emphasize that the large optical anisotropy is directly associated with open-ended nodal-lines in NbAs$_2$ and the flat spectral response ((Fig. \ref{fig3}(a) inset) is distinct from other anisotropic systems \cite{nakajima2011,frenzel2017}.

Having established the zero-field signatures of nodal-line fermions, we proceed to explore the properties of these anisotropic Dirac quasiparticles through magneto-optics. The electromagnetic signature of massive Dirac systems is the $\sqrt{B}$-spaced LLs dispersing from the gap energy $2\Delta$ \cite{chen2017,shao2017}, in contrast to $\sim\!B$-spaced LLs for parabolic bands \cite{shao2017}. The Dirac dispersion perpendicular to the nodal-lines can therefore be identified from magneto-optics. Unpolarized light (Fig. \ref{fig4}(a)) was used for magneto-reflectance measurement up to 17.5 T. A series of peaks (labeled 0 to 3) harden with increasing $B$ field and a weakly-dispersing feature shows up at lower energy ($\sim$85 meV). Both series of peaks start dispersing at finite energies, in sharp contrast to massless Dirac fermions \cite{chen2015b,yuan2017}.

Noticing the remarkable agreement of the step energies (Fig. \ref{fig3}(a)) and low-field peak energies (Fig. \ref{fig4}), we attribute the peaks in R(B)/R(0 T) to the interband LL transitions across the gapped Dirac bands. In Fig. \ref{fig4}(b), we plot the derivative contour $d$R/$d$B, which emphasizes the peaks in R(B)/R(0 T) as zero derivative (white) region bounded by positive (red) and negative (blue) derivative. The derivative plot is extremely sensitive to weak features in R(B)/R(0 T) and has been successfully used to investigate the subtle but important features in TI surface states \cite{schafgans2012}. 

We obtained $d$R/$d$B contours for both E$\,||\,b$ and E$\,||\,a$ polarizations and they show very similar features associated with peaks above the gap 2$\Delta_2$. The features related to the smaller gap 2$\Delta_1$ are only present for E$\,||\,b$, while completely suppressed for E$\,||\,a$. Interestingly, while $\sigma_1^{a}$ is smaller than $\sigma_1^{b}$ (Fig. \ref{fig2}(b)), the amplitude of the R(B)/R(0 T) is larger for E$\,||\,a$ than for E$\,||\,b$ (Fig. S7). Furthermore, the $d$R/$d$B plot for E$\,||\,b$ polarization (Fig. \ref{fig4}(b)) shows prominent dispersing features (grey dashed lines) starting $\sim$95 meV, in between the two gap energies (2$\Delta_1$ and 2$\Delta_2$).  This finite intercept at $B\rightarrow$ 0 T is anomalous and the exact nature of these resonances is a subject of future studies. An intriguing possibility pertains to the predicted topological surface states \cite{xu2016a,gresch2017,luo2016}, suggesting clear surface states for E$\,||\,b$ but not for E$\,||\,a$ (Fig. S8).

Besides uncovering weak dispersing features, the derivative plot $d$R/$d$B directly visualizes the $\sqrt{B}$ dispersion of LL transitions in NbAs$_2$. For massive Dirac nodal-lines, we have derived the following LL spectrum with perpendicular magnetic field $B$ (see SI Sec. III.):
\begin{equation}\label{eq2}
E_{\pm n} = \pm\sqrt{2 e\hbar |n| B v_1 v_2 \mathrm{cos}(\phi) + \Delta^2}
\end{equation} 
where $n$ is the LL index and $\phi$ is the angle between local nodal-line direction and magnetic field. $\Delta$ is the half-gap that characterizes the mass of the Dirac fermions $m_{x,y,z}^D = \Delta/v_{x,y,z}^2$ and the $\pm$ selects the conduction/valence band LLs. The dipole selection rules for interband LL transitions are $\delta |n|=|n|^{'}$-$|n|=\pm 1$. The transition energy is therefore:
\begin{equation} \label{eq3}
    E_T = \sqrt{2e\hbar |n|B \overline{v}^2 + \Delta^2} + \sqrt{2e\hbar (|n| + 1)B \overline{v}^2 + \Delta^2}
\end{equation}
where the effective velocity $\overline{v}=\sqrt{v_1 v_2 \mathrm{cos}(\phi)}$.

In Figs. \ref{fig4}(b),(c), green dashed lines are fitted interband LL transitions using Eq. \ref{eq3} with $\overline{v}\!=\!2.53\,\text{eV}\!\cdot\!\text{\AA}$ and gap $2\Delta_2$ = 114 meV. The fitted effective velocity are very close to the theoretical calculation (2.3 $\text{eV}\!\cdot\!\text{\AA}$) using the same asymptotic velocities $v_1$, $v_2$ for calculating $\sigma_1(\omega)$ (SI Sec. II.C).
Green dots are peak energies extracted from Fig. \ref{fig4}(a), showing excellent agreement for $\sqrt{B}$-spaced interband LL transition across $2\Delta_2$ in both unpolarized and polarized data. The non-linearly spaced LLs can also be easily identified at fixed $B$ as higher order LLs become closer in energy, in stark contrast to parabolic bands. The Dirac mass $m_{ab}^D=\Delta_2/v_a v_b=0.068\,m_e$ is $\sim$4 times smaller compared to the trivial carriers (0.24-0.29 $m_e$) \cite{shen2016,yuan2016a}. This much smaller mass implies that the high mobility carriers in NbAs$_2$ could come from Dirac fermions in the nodal-lines.

The extracted gap energy ($2\Delta_2\!\sim$114 meV) from fitting the LL dispersion is very close to the step energy in the zero-field data ($2\Delta_2$+$2E_{F2}\!\sim$120 meV), indicating that the gapped cones are only weakly doped ($E_F<5$ meV). This low doping-level in the massive nodal-lines (near X$_1$-Y) gives rise to a huge magneto-infrared response (Fig. S7) since it can be easily driven into the extreme quantum limit (only 0th LL occupied). In contrast, the heavy trivial bands with large carrier density remain in the classical region at the highest measured field (17.5 T) \cite{shen2016,yuan2016a}.

We now discuss the implication of massive Dirac nodal-lines for the unusual magneto-transport properties of NbAs$_2$. Gigantic MR ($>10^5\,\%$) in nonmagnetic NbAs$_2$ has been observed \cite{shen2016,yuan2016a,li2016a,wang2016c} and explained as a cooperation of perfect electron-hole compensation and high-mobility carriers, leading to a $\sim\!B^2$ increase. However, high-field MR measurements clearly deviate from the $B^2$ dependence starting at $\sim$10 T and linearly increase with $B$ without saturation \cite{yuan2016a}. Such large ($>10^5\,\%$), non-saturating and crossover from (nearly) quadratic- to linear-increasing behavior calls for interpretations beyond electron-hole compensation. We believe the lightly-doped Dirac nodal-lines established here are crucial to understand the unusual MR in NbAs$_2$. 

The quantum linear MR \cite{abrikosov1998} reads $\rho_{xx}\!=\!N_{i}B/\pi n^2 e$, where $N_i$ is the scattering center concentration and $n$ is the carrier density. Both the minority massive Dirac fermions in the quantum limit and the majority carriers in the classical two-band model can give rise to large MR. However, a slight deviation from perfect electron-hole compensation, which exists in NbAs$_2$, will cause the $\sim\!B^2$ increasing MR to saturate at a field-independent value \cite{ali2014}, contrary to experiment \cite{yuan2016a}. The existence of massive Dirac fermions naturally explains the discrepancies: when the classical MR saturates at an intermediate field, the quantum linear MR from Dirac fermions dominates and hence the dependence changes.

In summary, we discover Dirac nodal-lines in NbAs$_2$ and establish theoretically the novel response functions for general dispersive nodal-lines. Our observation not only sheds light on the interpretation of the exotic MR in this family of materials, but also paves the way for identifying new NLSMs using optical/magneto-optical spectroscopy. NbAs$_2$ therefore constitutes a concrete platform to realize various predictions for massive nodal-line fermions, including large spin Hall effect \cite{sun2017} and Floquet Weyl points \cite{yan2016a,narayan2016,chan2016a}. 

\begin{acknowledgments}
This research is supported by ARO grant W911nf-17-1-0543. D.N.B. is the Moore Foundation Investigator, EPIQS Initiative Grant GBMF4533. F.C.C acknowledges the support provided by the Ministry of Science and Technology in Taiwan under Grant No. 106-2119-M-002-035-MY3. A portion of this work was performed at the National High Magnetic Field Laboratory, which is supported by the National Science Foundation Cooperative Agreement No. DMR-1157490 and the State of Florida. A.J.B. and M.B.M were supported by the
US Department of Energy, Office of Basic Energy Sciences, Division of Materials Sciences and Engineering, under Grant
No. DEFG02-04-ER46105.
\end{acknowledgments}

\end{document}